\documentclass[pra,superscriptaddress,amsfonts,amsmath,floatfix,twocolumn,showpacs]{revtex4}
\usepackage{graphicx}
\usepackage{epstopdf}
\usepackage{amssymb}
\usepackage{amsmath}
\usepackage{xspace}
\usepackage{verbatim}
\usepackage{color}
\newcommand{\bra}[1]{\left\langle #1 \right|}
\newcommand{\ket}[1]{\left| #1 \right\rangle}

\begin{document}
\title{Nonlinearity-Induced Entanglement Stability in a Qubit-Oscillator System}

\author{V\'ictor Montenegro}
\email{v.montenegro.11@ucl.ac.uk}
\affiliation{Department of Physics and Astronomy, University College London, Gower Street, London WC1E 6BT, United Kingdom}

\author{Alessandro Ferraro}
\email{a.ferraro@qub.ac.uk}
\affiliation{Centre for Theoretical Atomic, Molecular and Optical Physics, School of Mathematics and Physics, Queen's University, Belfast BT7 1NN, United Kingdom}

\author{Sougato Bose}
\email{sougato@theory.phys.ucl.ac.uk}
\affiliation{Department of Physics and Astronomy, University College London, Gower Street, London WC1E 6BT, United Kingdom}

\date{\today}

\begin{abstract}
We consider a system composed of a qubit interacting with a quartic (undriven) nonlinear oscillator (NLO) through a conditional displacement Hamiltonian. We show that even a modest nonlinearity can enhance and stabilize the quantum entanglement dynamically generated between the qubit and the NLO. In contrast to the linear case --- in which the entanglement is known to oscillate periodically between zero and its maximal value --- the nonlinearity suppresses the dynamical decay of the entanglement once it is established. While the entanglement generation is due to the conditional displacements, as noted in several works before, the suppression of its decay is related to the presence of squeezing and other complex processes induced by two- and four-phonon interactions. Finally, we have solved the respective Markovian master equation, showing that the previous features are preserved also when the system is open.
\end{abstract}


\maketitle

\section{Introduction} 
Two-level quantum systems (qubits) and quantum harmonic oscillators are the two most basic building blocks in quantum information science. Stimulated by this, in the last decades there have been remarkable experimental progresses in the accurate control of the interaction in qubit-oscillator systems, including: trapped ions \cite{IONS}, cavity-QED \cite{CQED} , ultracold atomic Bose-Einstein condensate \cite{TREUTLEIN, HUNGER}, quantum dots or Cooper-pair boxes \cite{HENNESY, STEELE, ARMOUR, BOSE}, superconducting qubits coupled to superconducting resonators \cite{CHIORESCU, WALLRAFF, CLARKE}, optomechanical systems \cite{SCALA, LI}, etc. Furthermore, they have been investigated in different qubit-oscillator coupling regimes, including the recently so-called ultrastrong regime, where the qubit-oscillator coupling strength is comparable to the qubit and oscillator energy scales \cite{ARNE, ASHHAB, HAUSINGER, TIAN, BOURASSA}.

In general, the quantum oscillator is modeled harmonically, however this is typically an approximation of more complicated scenarios. In fact, quantum nonlinear oscillators (NLO) have been implemented in several settings, including trapped ions (where the trapping potential can be modified to include nonlinearities \cite{HOME}), optomechanical systems (where tunable nonlinearities have been realized \cite{SANKEY}), and atoms in optical lattices \cite{LEWENSTEIN}. Interestingly, it has been shown that the inclusion of strong enough nonlinearities in the oscillator potential allows new possibilities to generate non-classical states \cite{DIVINCENZO, ONG, PEANO, KOLKIRAN, ANDERSSON, RIPS}. However, despite the promising experimental progresses in the control and fabrication of NLO, it is still a challenge to achieve significant nonlinearities (for a more detailed discussion about the nonlinear regimes and their possible experimental implementations see Sec. \ref{Anharmonic Potential}). Remarkably, we will 
show here that also weak non-linearities suffice to provide non-trivial and potentially useful features in the context of a qubit-NLO setting.

In this article we consider a quantum system composed of a qubit interacting with a quartic (undriven) NLO through a conditional displacement Hamiltonian. In order to contrast our results when the nonlinearity is included in the potential, we first solve the simplest case, i.e., a qubit interacting with a quantum harmonic oscillator. In this case, the entanglement is generated periodically as a consequence of the superposition principle. First, by including a weak nonlinear perturbation in the oscillator potential, we have obtained analytically the wave function in the rotating-wave approximation --- in a regime where both the qubit-NLO coupling as well as the nonlinearity strength are small compared to the oscillator frequency. In this case, an explicit Kerr-like term in the evolution appears, generating quadrature squeezing for short times. In particular, we will show that the entanglement generated in this nonlinear scenario is larger with respect to the linear case and, in addition, it dynamically 
reaches a stabilization region. For very large times the oscillator shows an intricate behavior exhibiting negative values in the Wigner distribution. 

The second main result of this article is obtained by taking into account a strong qubit-NLO coupling, while still considering a weak nonlinear regime. The novelty with respect to previous works is the inclusion of the two- and four-phonon processes, i.e. the full numerical dynamics of the system without any approximation. In this case \textit{i)} the entanglement stabilization region is achieved faster than in the weak qubit-NLO coupling case, and \textit{ii)} the entanglement reaches its maximum value. Finally, we have solved the Markovian master equation, taking into account only the damping of the oscillator, and even in this case the system dynamics remains robust showing the main features just described for a considerable number of cycles.

The article is organized as follow. In section \ref{Non-dissipative dynamics} we present the system under consideration for the non-dissipative case. In section \ref{Without nonlinearity: Exact solution of the wave-function} we solve the system in the linear case for the sake of comparison with the results presented in section \ref{Anharmonic Potential} where we consider the full nonlinear dynamics. We focus on two regimes: in subsection \ref{WeakCoupling} we consider the weak qubit-NLO coupling regime, in which an analytical approximation can be obtained for the dynamics of the system wave-function; in subsection \ref{StrongCoupling} we show the full numerical solution for the strong coupling regime. Furthermore, we present the case when losses are present in the system and suggest possible experimental implementations. Finally, we give some concluding remarks in section \ref{Concluding Remarks}.

\section{THE MODEL}\label{Non-dissipative dynamics}

We consider a two-level system (qubit) coupled to a quartic nonlinear oscillator, as described by the Hamiltonian
\begin{equation}
 \hat{H} = \hat{H}_q + \hat{H}_{o} + \hat{H}_{q-o}\label{Hinicial}\;,
\end{equation}
where $\hat{H}_q$ ($\hat{H}_{o}$) is the free qubit (NLO) Hamiltonian and $\hat{H}_{q-o}$ is their mutual interaction. Each term above is defined as follows
\begin{eqnarray}
\hat{H}_q &=& \hbar\omega_q\hat{\sigma}_z,\label{ha}\\
\hat{H}_{o} &=& \frac{1}{2m}\hat{p}^2 + \frac{1}{2}m\omega_o^2\hat{x}^2 + \tilde{\delta}\hat{x}^4,\label{hm}\\
\hat{H}_{q-o} &=& - \hbar \tilde{g}\hat{\sigma}_z\hat{x},\label{ham}
\end{eqnarray} 
where $\hbar\omega_q$ corresponds to the qubit separation energy between its ground $(\ket{\downarrow})$ and excited ($\ket{\uparrow}$) states, $\hat{\sigma}_z$ is the usual Pauli z---(pseudo)spin matrix ($\hat{\sigma}_z\ket{\uparrow} = \ket{\uparrow}, \hat{\sigma}_z\ket{\downarrow} = -\ket{\downarrow}$), $\omega_o$ is the frequency of the oscillator in absence of nonlinearities, $\tilde{\delta}$ is the quartic nonlinear strength, whereas $\hat{x}$ and $\hat{p}$ are the usual position and momentum operators, respectively. In Eq. (\ref{ham}) the interaction strength is parametrized by $\tilde{g}$ (assumed to be positive throughout) and it is linear in the position operator $\hat{x}$. This type of interaction has been realized/proposed in various experimental settings --- including ion traps \cite{WUNDERLICH}, cavity-QED \cite{SOLANO}, and nanomechanical resonators \cite{RABL} --- and its action can be understood as a displacement of the oscillator conditioned on the state of the qubit. As such, it has 
been exploited for example as a tool for reconstructing the state of quantum oscillators in various physical systems \cite{TUFARELLI, TUFARELLI2}, or as a mediator to induce qubit-qubit interactions \cite{WUNDERLICH}.

Let us notice that in order for the Hamiltonian (\ref{hm}) to be valid, in the following we will consider a modest nonlinear quartic perturbation. In particular, we require that $\tilde{\delta}\langle \hat{N} \rangle \ll \omega_o$ during the evolution (where $\langle \hat{N} \rangle$ is the average phonon number for the oscillator), thus ensuring that the single-frequency assumption for the oscillator ($\omega_o$) remains valid.

Introducing the usual annihilation $\hat{a}$ and creation $\hat{a}^\dag$ operators for the oscillator we can recast the oscillator canonical operators as 
\begin{eqnarray}
 \hat{x} &=& \sqrt{\frac{\hbar}{2m\omega_o}}(\hat{a}^\dag + \hat{a}) \label{x}\\
 \hat{p} &=& i\sqrt{\frac{m\hbar\omega_o}{2}}(\hat{a}^\dag - \hat{a}) \label{p}\;,
\end{eqnarray}
Rescaling Eq.~(\ref{Hinicial}) by $\hbar\omega_o$ and switching to the interaction picture with respect to the qubit, the relevant Hamiltonian reads
\begin{eqnarray}
 \hat{H}_{int} = \hat{a}^\dag\hat{a} + \delta(\hat{a}^\dag + \hat{a})^4 - k \hat{\sigma}_z(\hat{a}^{\dag} + \hat{a}),\label{han}
\end{eqnarray}
where,
\begin{eqnarray}
g &=& \tilde{g}\sqrt{\frac{\hbar}{2m\omega_o}},\\
\delta &=& \frac{\tilde{\delta}}{\hbar\omega_o}\left(\frac{\hbar}{2m\omega_o}\right)^2,\\
k &=& \frac{g}{\omega_o}.
\end{eqnarray}

In general, throughout this work we will consider the following initial state:
\begin{equation}
 \ket{\psi(0)} = \frac{1}{\sqrt{2}}(\ket{\uparrow} + \ket{\downarrow})\otimes\ket{\alpha},\label{psi0}
\end{equation}
where the oscillator coherent state is defined as $\ket{\alpha} = \mathrm{exp}\left[\alpha\hat{a}^\dag - \alpha^*\hat{a}\right]\ket{0} = \hat{D}(\alpha)\ket{0}$ ($\hat{D}(\alpha)$ is the usual displacement operator).

\section{dynamics in absence of nonlinearity}\label{Without nonlinearity: Exact solution of the wave-function}
\begin{center}
\begin{figure}
\includegraphics[scale=0.45]{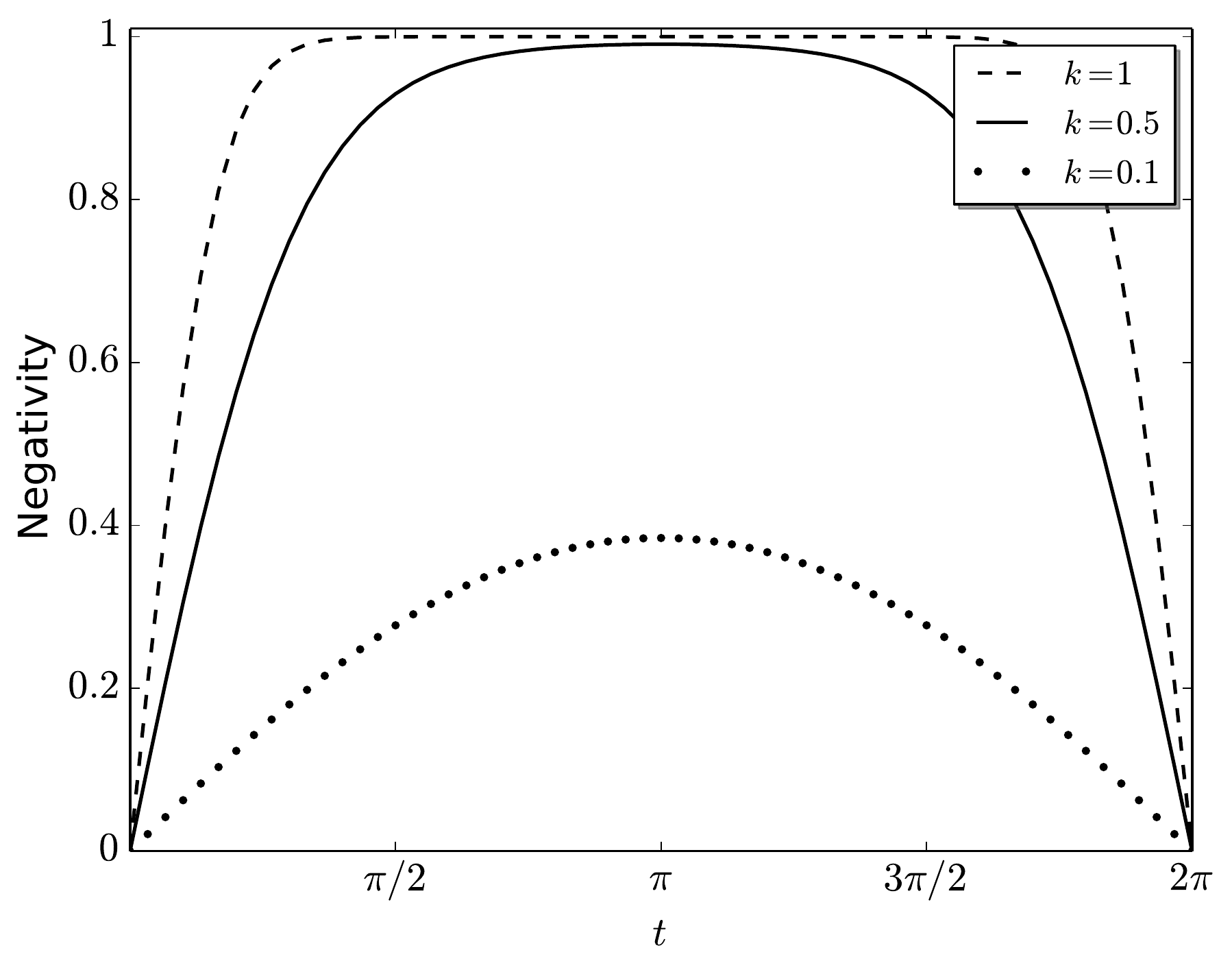}
\caption{(color online) Time dependence of the Entanglement negativity for different values of $k$ in absence of nonlinearities in the NLO potential. Starting from the separated state given by Eq. (\ref{psi0}) ($\alpha = 2$) the system becomes entangled ($0 < t < 2\pi$), reaching a maximum at $t = \pi$. Finally, at $t = 2\pi$ the system return to its original state, thus the negativity is zero. In this and all the figures, $t$ is a scaled time, corresponding to the actual time multiplied by $\omega_o$.}
\label{N_d_0_kvar}
\end{figure}
\end{center}

\begin{center}
\begin{figure}
\centering
\includegraphics[scale = 0.45]{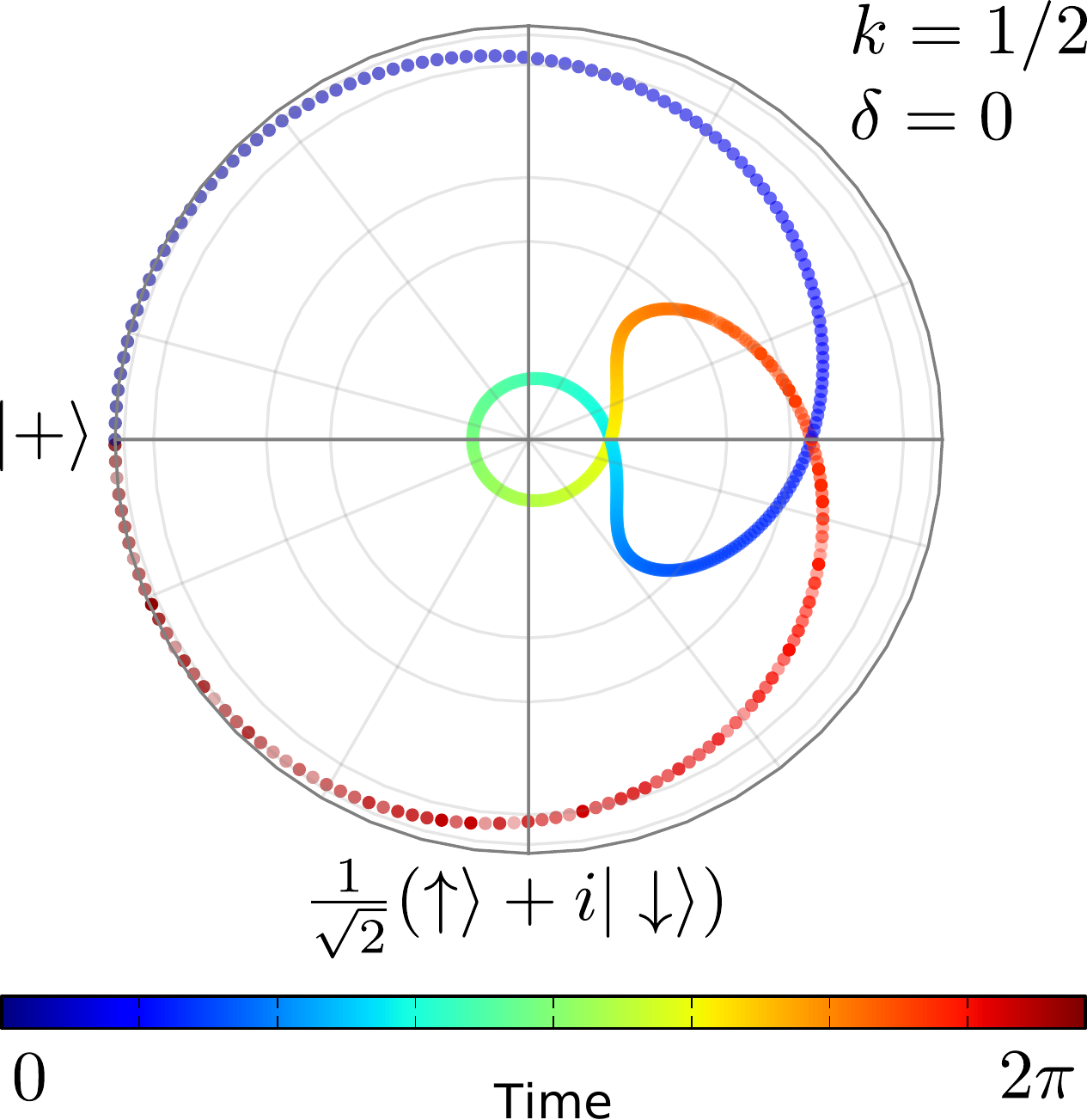}
\caption{(color online) Dynamics of the reduced density operator for the qubit state in the Bloch Sphere (top-view) with $k = 0.5, \alpha = 2, \delta = 0$. We can see that in absence of nonlinearities the qubit dynamics remains periodically for the whole evolution. Here, the leftmost (lowermost) point of the $x-$axis ($y-$axis) represents the state $\ket{+}(\frac{1}{\sqrt{2}}(\ket{\uparrow}) + i\ket{\downarrow})$. }\label{Bloch_Sphere}
\end{figure}
\end{center}

\begin{center}
\begin{figure}
  \includegraphics[scale = 0.45]{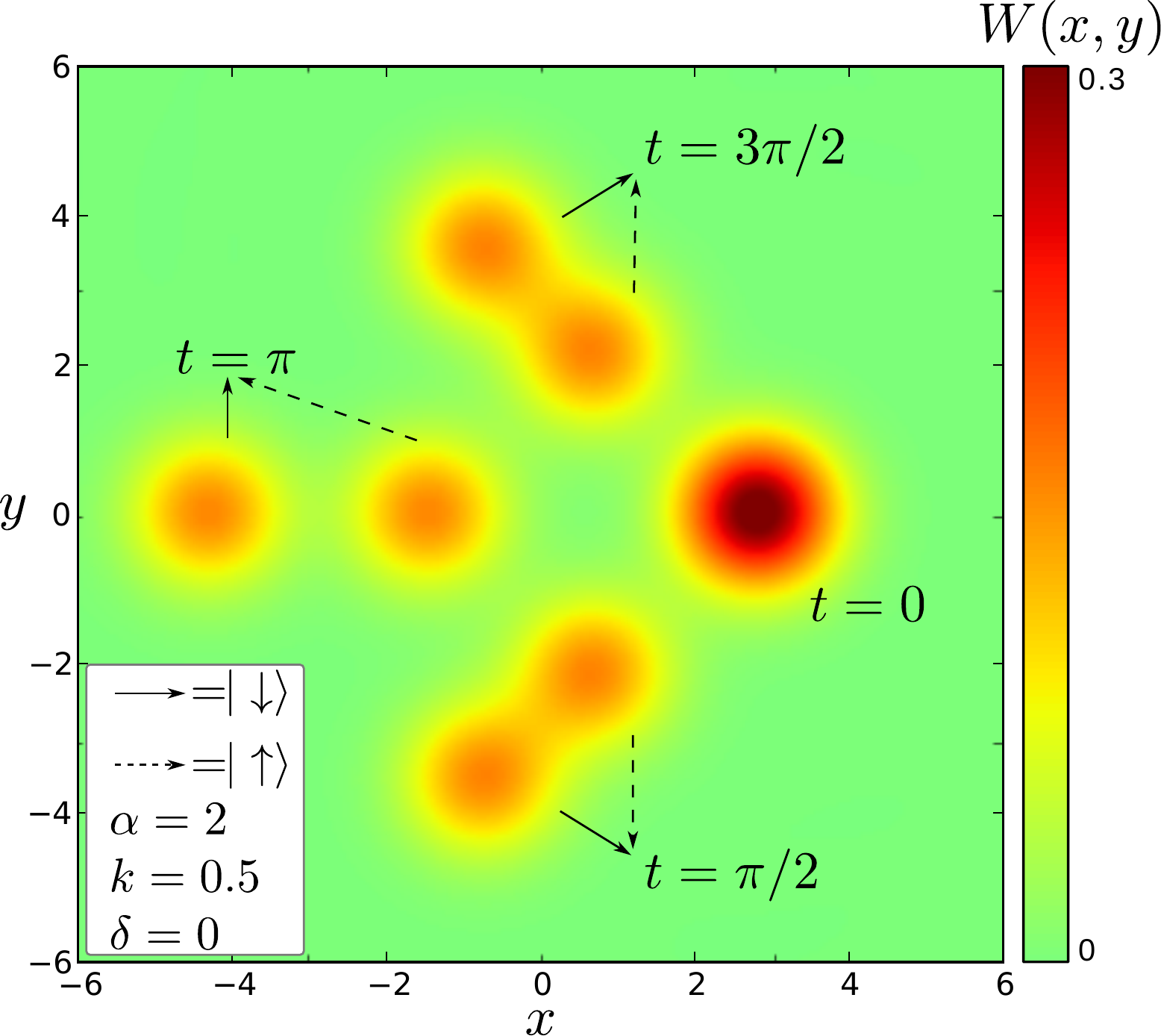}
  \caption{(color online)  The figure shows the Wigner function $W(x,y)$ of the reduced density operator for the NLO associated with the Eq.~(\ref{negdelta0}). The single peak of the initial coherent states separates into two components, each associated with a different qubit eigenstate. Specifically, the solid line arrow (dashed line arrow) indicates the component associated with $\ket{\downarrow} (\ket{\uparrow})$. The Wigner function is defined as $W(x,y) = \frac{1}{\pi\hbar}\int_{-\infty}^\infty\left<x + x'\left| \hat{\rho}_{osc}\right| x-x' \right>e^{-2iyx'/\hbar}dx'$ and the axis are given accordingly.}\label{wigner_cycle}
\end{figure}
\end{center}

For the sake of comparison with the genuine features of an anharmonic oscillator, we briefly summarize here the results for the case of a simple quantum harmonic oscillator [$\delta = 0$ in Eq. (\ref{han})]. It is straightforward to obtain the time dependent solution for this system (see Appendix \ref{appA})
\begin{equation}
\ket{\psi(t)} = \frac{1}{\sqrt{2}}(e^\Phi\ket{\uparrow}\otimes\ket{\alpha_{\uparrow}} + e^{-\Phi}\ket{\downarrow}\otimes \ket{\alpha_{\downarrow}}) \label{negdelta0}
\end{equation}
where
\begin{eqnarray}
 \Phi &=& ik \mathbb{I}\mathrm{m}\left[\alpha\eta\right] = ik\alpha \mathrm{sin}(t),\\
 \ket{\alpha_{\uparrow}} &=& \ket{\alpha e^{-it} + k\eta},\\
 \ket{\alpha_{\downarrow}} &=& \ket{\alpha e^{-it} - k\eta}
\end{eqnarray}
with  $(\eta = 1-\mathrm{exp}\left[-it\right])$. The above solution implies that the wave function is periodic and, in particular, the initial separable state is recovered at times $2\pi n$, $n$ being an integer. On the other hand, for $0 < t < 2\pi$, the oscillator is entangled with the qubit --- this hybrid entanglement reaching its maximum at time $t = \pi$. In order to quantify the entanglement we use the negativity \cite{NEGATIVITY1, NEGATIVITY2} (as we will eventually also compute the same entanglement when the oscillator is an open system, and negativity is a measure also valid for that case). The negativity can be computed as
\begin{equation}
 N(t) = \frac{1}{2}\sum_i(|\lambda_i| - \lambda_i),
\end{equation}

where the $\lambda_i$ are the eigenvalues of the partially transposed qubit-NLO density matrix at fixed time $t$. The time dependence is shown in Fig.~\ref{N_d_0_kvar} for a fixed coherent state ($\alpha = 2$) and different couplings $k$. A similar dynamics has been reported in analogous optomechanical settings (see \textit{e.g.} Ref.~\cite{BOSE2}). It is of relevance to notice at this stage that the entanglement generated so far is due only to the interlinked dynamics of the qubit and the oscillator, as generated by the conditional displacement of Eq.~(
\ref{ham}). In the next section we will add another type of entanglement source to the system when 
a nonlinearity is added.
The periodicity of the system can be also appreciated from the reduced density matrix for the qubit $\hat{\rho}_q = \mathrm{Tr}_{osc}(\ket{\psi(t)}\bra{\psi(t)})$:
\begin{equation}
\hat{\rho}_q = \frac{1}{2}\left[\ket{\uparrow}\bra{\uparrow} + \mathrm{e}^{4k^2(\mathrm{cos}(t) - 1)}(\ket{\uparrow}\bra{\downarrow} + \ket{\downarrow}\bra{\uparrow}) + \ket{\downarrow}\bra{\downarrow}\right]\label{qubit}
\end{equation}
given that $\langle
\hat{\sigma}_z \rangle = 0$, we can easily plot in Fig.~\ref{Bloch_Sphere} the Bloch sphere top-view of the Bloch vector of $\hat{\rho}_q$. 
Another feature immediately evident from the solution in Eq.~(\ref{negdelta0}) is that the dynamics of each qubit eigenstate is linked to that of a coherent state during the evolution (\textit{e.g.}, the eigenstate $\ket{\uparrow}$ is linked to $\ket{\alpha e^{-it} + k\eta}$). In order to better appreciate this behavior, as well as the oscillator dynamics, we have calculated the Wigner function of the reduced density operator for the oscillator  $\hat{\rho}_{osc} = \mathrm{Tr}_{q}\left[\ket{\psi(t)}\bra{\psi(t)}\right]$. In Fig.~\ref{wigner_cycle} we plot the Wigner function of the reduced density operator for the NLO associated with the Eq.~(\ref{negdelta0}). As we can see, if the initial state is $\ket{\downarrow,\alpha}$ the oscillator's Wigner function rotates in a larger circle with respect to the $\ket{\uparrow,\alpha}$ initial state.

\section{Nonlinear dynamics}\label{Anharmonic Potential}
We will now derive the central results of this work. In particular, in Sec.~\ref{WeakCoupling} we study the evolution for the weak coupling regime ($\{k,\delta\} \ll 1$), where an approximated analytical expression for the wave-function can be obtained. In Sec.~\ref{StrongCoupling} we present the general results in the strong coupling regime ($i.e., k \approx 1, \delta \ll 1$), considering as well the detrimental effects of noise.

\subsection{Weak qubit-NLO coupling regime : Approximated analytical solution for k $\ll$ 1, $\delta \ll 1$}\label{WeakCoupling}
We will refer to the weak coupling regime when the rescaled qubit-NLO coupling strength is much lower than the qubit and oscillator free energies. In order to investigate the perturbation in the NLO we rewrite the quartic term as follows
\begin{equation}
 (\hat{a}^\dag + \hat{a})^4 = \hat{\mathcal{A}}_4 + \hat{\mathcal{A}}_2 + \hat{\mathcal{A}}_{ns},\label{a4}
\end{equation}
where we have emphasized the phonon process contributions; namely, $ \hat{\mathcal{A}}_{i=2,4,ns}$ correspond to the operators identifying two- and four-phonon transitions and the number-state contribution $(ns)$, respectively. Considering the commutation rule $[\hat{a},\hat{a}^\dag] = 1$ one obtains
\begin{eqnarray}
 \hat{\mathcal{A}}_{4}  &=& \hat{a}^{\dag 4} + \hat{a}^4,\\
 \hat{\mathcal{A}}_{2}  &=& 6(\hat{a}^{\dag 2} + \hat{a}^2) + 4(\hat{a}^{\dag 2}\hat{a}^\dag\hat{a} + \hat{a}^\dag\hat{a}\hat{a}^2),\\
 \hat{\mathcal{A}}_{ns} &=& 6((\hat{a}^\dag\hat{a})^2 + \hat{a}^\dag\hat{a}),
\end{eqnarray}
and the Hamiltonian in Eq. (\ref{han}) reads as
\begin{eqnarray}
 \nonumber \hat{H} &=& (1 + 6\delta)\hat{a}^\dag\hat{a} + 6\delta(\hat{a}^\dag\hat{a})^2 - k \hat{\sigma}_z(\hat{a}^{\dag} + \hat{a}) \\ &+& \delta(\hat{\mathcal{A}}_2 + \hat{\mathcal{A}}_4).\label{happ}
\end{eqnarray}
In the equation above, the terms in the second line correspond to two- and four-phonon transitions and they can both be neglected by invoking a rotating wave approximation. By considering a frame rotating with the free oscillator Hamiltonian, one can recast Eq.~(\ref{happ}) as 
\begin{eqnarray}
 \nonumber \hat{H}_{\rm int} &=&  6\delta \hat{a}^\dag\hat{a} + 6\delta(\hat{a}^\dag\hat{a})^2 - k \hat{\sigma}_z\left(\hat{a}^{\dag} e^{+it} + \hat{a}e^{-it}\right) \\ &+& \delta \left(
6e^{+2it}\hat{a}^{\dag^2}
+4e^{+2it}\hat{a}^{\dag^2}\hat{a}^\dag\hat{a} 
+e^{+4it}\hat{a}^{\dag^4} 
+H.c.\right) \nonumber \\
\label{happ2}
\end{eqnarray}
Among the terms proportional to the nonlinearity strength $\delta$, the oscillating ones can be approximately neglected. Thus, transforming back to the Schr\"{o}dinger picture, one has the following Hamiltonian:
\begin{equation}
 \hat{H}_{\rm RWA} \approx (1 + 6\delta)\hat{a}^\dag\hat{a} + 6\delta(\hat{a}^\dag\hat{a})^2 - k\hat{\sigma}_z(\hat{a}^{\dag} + \hat{a})\label{ah_app}
\end{equation}

Using the same techniques as before (see Appendix \ref{appB}) we obtain the following solution for the wave function, where we have neglected the terms proportional to $\{k\delta, k^2\delta, k^3\delta\}$ (for simplicity we have considered real amplitudes for the coherent state):
\begin{eqnarray}
 \nonumber \ket{\psi(t)} &=& \frac{1}{\sqrt{2}}\hat{D}(+k)\mathrm{exp}\left[-6it\delta(\hat{a}^\dag\hat{a})^2\right]\ket{\uparrow}\otimes\ket{\tilde{\alpha}_{\uparrow}}\\
 &+& \frac{1}{\sqrt{2}}\hat{D}(-k)\mathrm{exp}\left[-6it\delta(\hat{a}^\dag\hat{a})^2\right]\ket{\downarrow}\otimes\ket{\tilde{\alpha}_{\downarrow}}\label{asmall}
\end{eqnarray}
where,
\begin{eqnarray}
 \ket{\tilde{\alpha}_{\uparrow}} &=&   \ket{e^{-i(1+6\delta)t}(\alpha - k)},\label{aae}\\
 \ket{\tilde{\alpha}_{\downarrow}} &=& \ket{e^{-i(1+6\delta)t}(\alpha + k)}.\label{aag}
\end{eqnarray}

\begin{center}
\begin{figure}
\includegraphics[scale = 0.45]{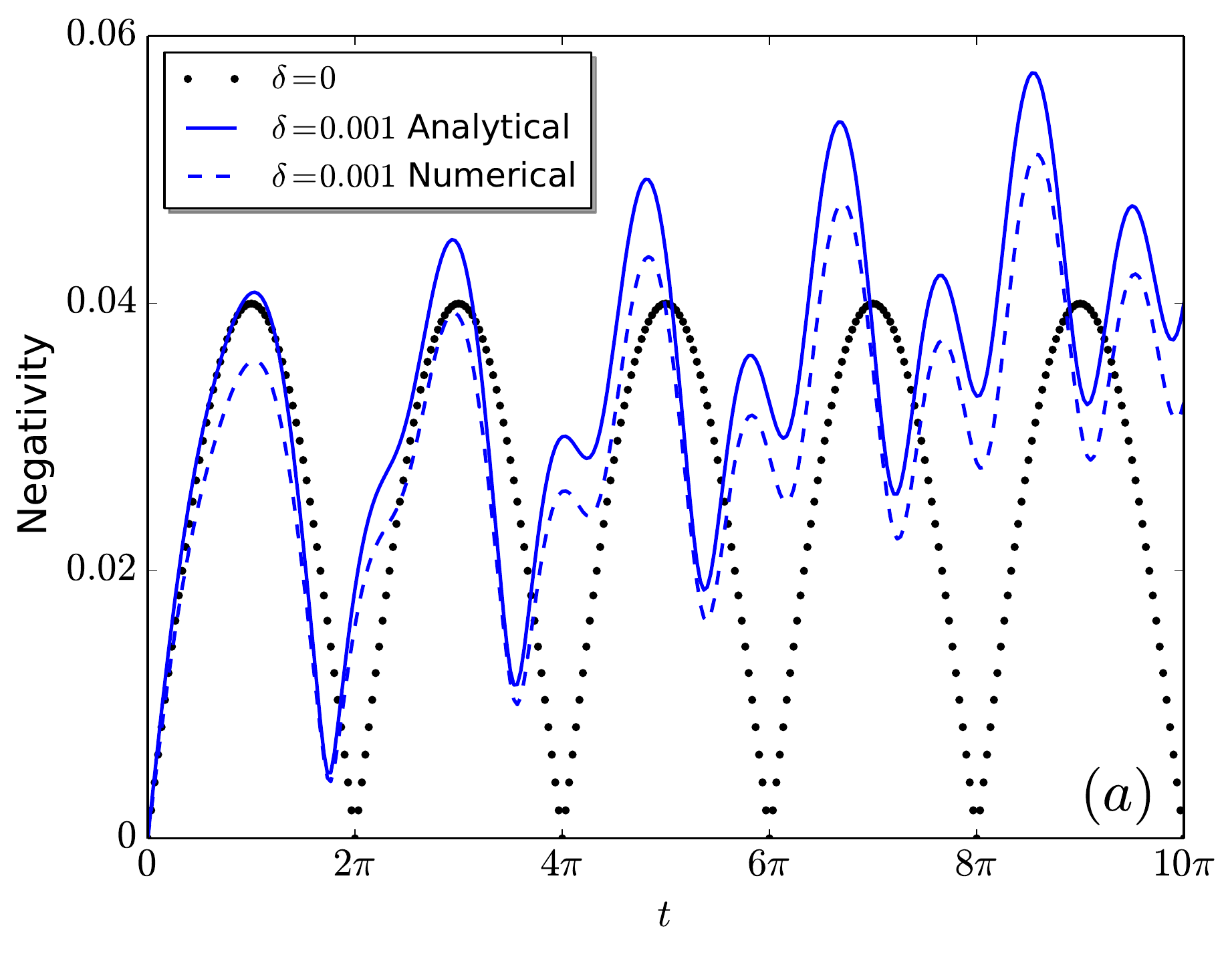}
\includegraphics[scale = 0.45]{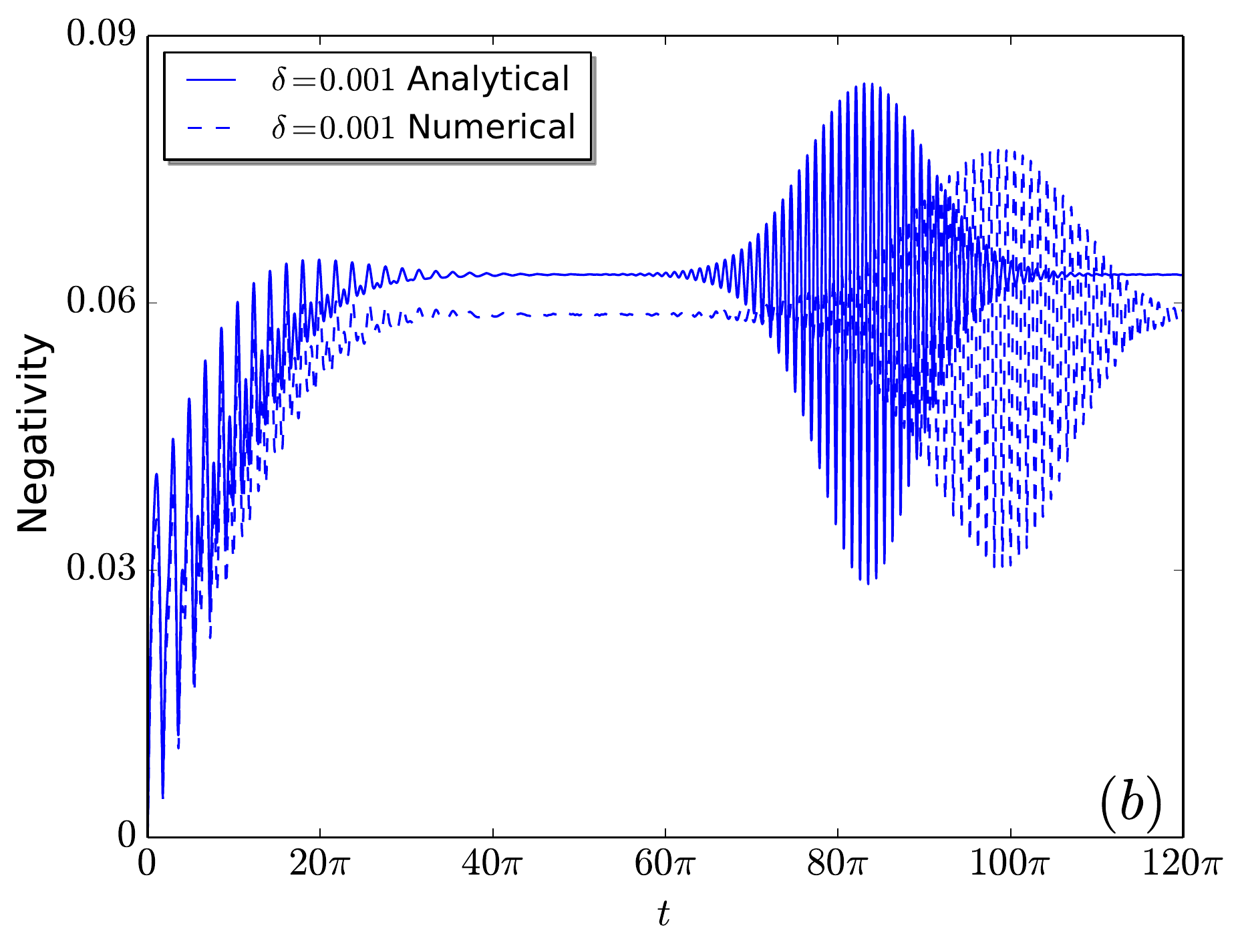}
\caption{(color online) (a) negativity as function of time $t$ for $k = 1/100$ and $\delta = 1/1000$ $(\alpha = 2)$. We compare the entanglement using an analytical expression (solid line) (Eq. (\ref{asmall})) and the numerical one (dashed line) using Eq. (\ref{happ}). The dotted line is the evolution in absence of nonlinearity. As we can see the inclusion of the nonlinear term increases the entanglement reaching a time-plateau or stabilization zone. (b) We compare the analytical expression with the numerical solution for the same set of parameters for larger times.}\label{n_small}
\end{figure}
\end{center}

A comparison between the approximate analytical results in Eq.~(\ref{asmall}) versus a numerical computation using the full original Hamiltonian in Eq.~(\ref{Hinicial}) is shown for short times in Fig.~\ref{n_small}-a, where we plot the negativity for $k=1/100$ and $\delta = 1/1000$ (for $\alpha = 2$). As we can see, the analytical approximation agrees reasonably well with the numerical results (the dotted line corresponds to the dynamics of the system in absence of nonlinearity). More importantly, the presence of a nonlinear Kerr-like term proportional to   $(\hat{a}^\dag\hat{a})^2$ represents a new source for entanglement and non-classical effects, allowing to grasp the main features associated with the full Hamiltonian. The first of these features is the lack of a periodic behavior for short times which implies, in particular, that the entanglement does not decrease to zero. In addition, the actual values of the negativity show a clear \textit{enhancement of the entanglement} with respect to linear case (
$\delta = 0$). Remarkably, as shown in Fig.~\ref{n_small}-b, after few cycles the negativity reaches a plateau, implying a \textit{time-stabilization of the entanglement} at values higher than the maximum attained for $\delta = 0$. For longer time-scales ($t \gg 120\pi$), the expected collapses and revivals appear only assuming both the rotating-wave approximation and small $k\delta$. On the other hand, the full numerical solution of the evolution does not show any collapse nor revival---in fact, the negativity never drops to zero. Due to the establishment of a stabilization window, we can define the width of the time plateau ($\Delta$) as the region in which the negativity does not show significant oscillations; e.g., in Fig.~\ref{n_small}-b a plateau is approximately achieved for $30\pi \leq t \leq 70\pi$, being its width $\Delta \approx 40\pi$. The dependence between $\Delta$ and $\delta$ eludes analytical calculations, however a straightforward numerical evaluation (under the constraints of $\{
k,\delta\} \ll 1$) shows the dependence to be inversely proportional to the nonlinearity strength---in fact, for $10^{-4} < k < 10^{-2}$ and $10^{-4} < \delta < 10^{-2}$, one can show that $\Delta \approx 0.1/\delta$.

\begin{center}
\begin{figure}
  \includegraphics[scale = 0.45]{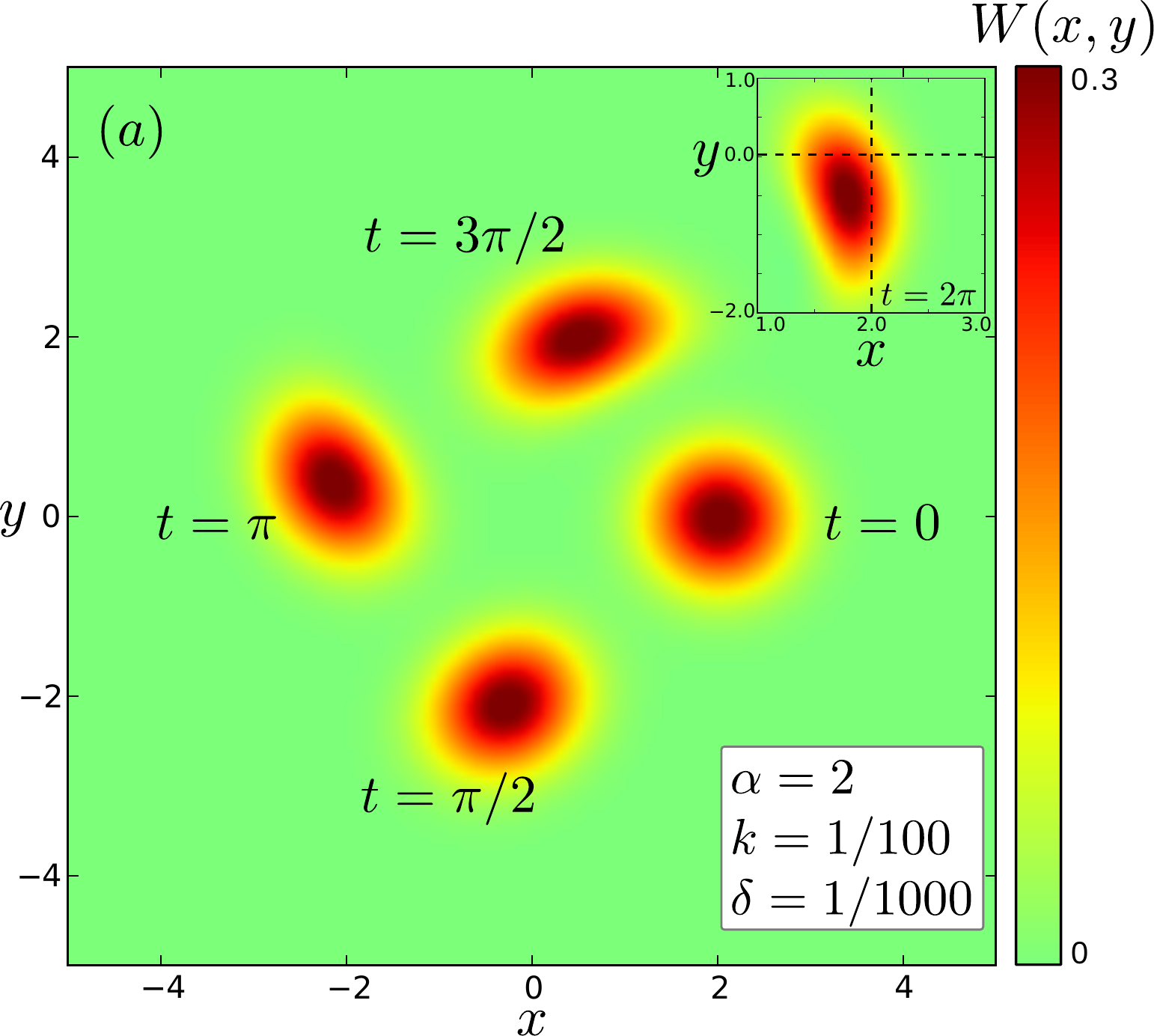}
  \includegraphics[scale = 0.45]{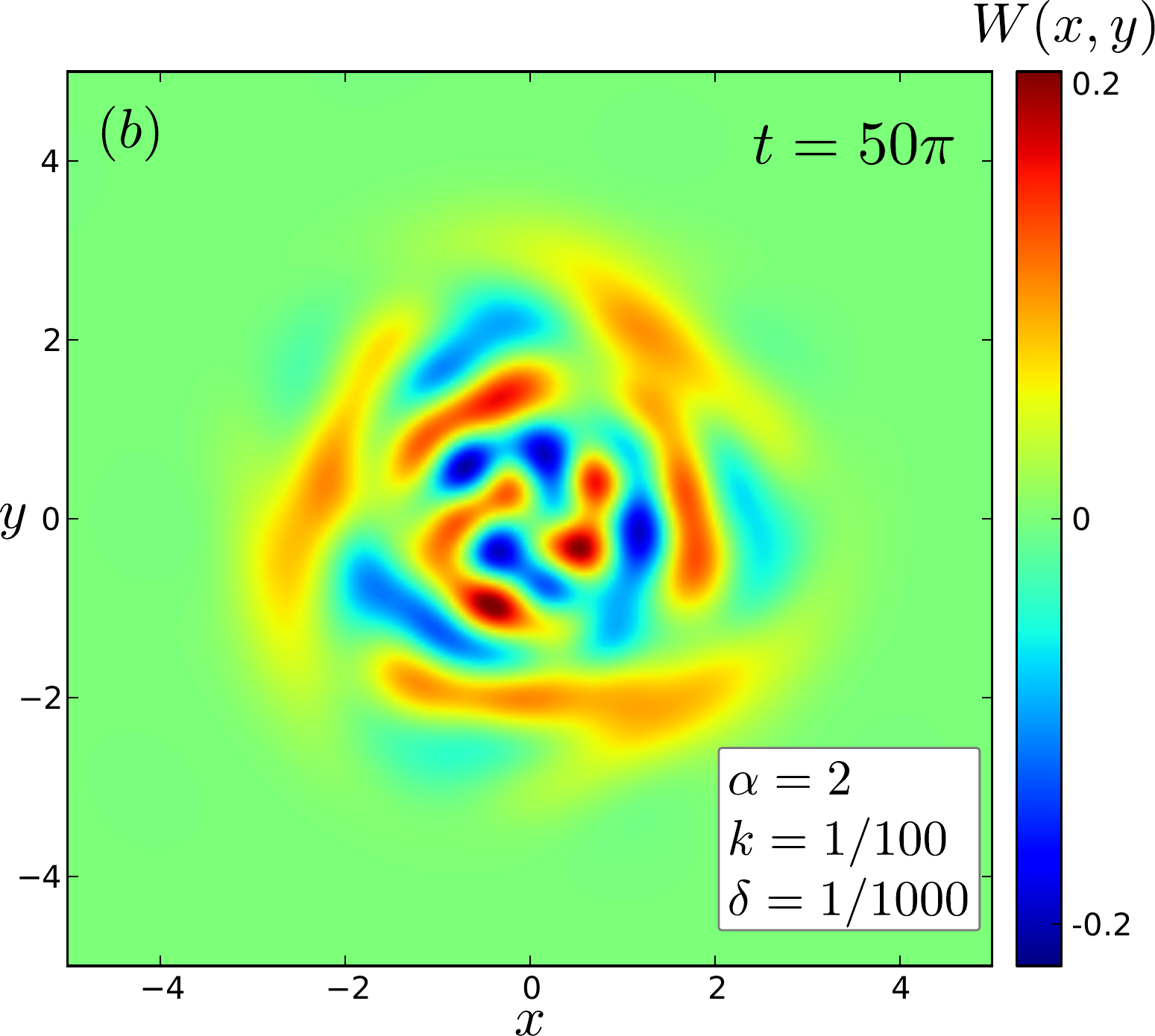}
  \caption{(color online) (a) The above figure shows the Wigner function $(W(x,y))$ for the oscillator state for $t=0,\pi/2,\pi,3\pi/2,2\pi$. The state considered here is the one in Eq. (\ref{asmall}). As we can see, in the first cycle, due to the small values of $k$ and $\delta$ both components of the qubit remains superposed during all time, showing squeezing in the quadratures $\{x,y\}$. (b) The below figure shows the state at $t=50\pi$, as we see the state becomes complex evidencing negatives values during the dynamics.}\label{wigner2D_small_k_delta}
\end{figure}
\end{center}

\begin{center}
\begin{figure}
  \includegraphics[scale=0.45]{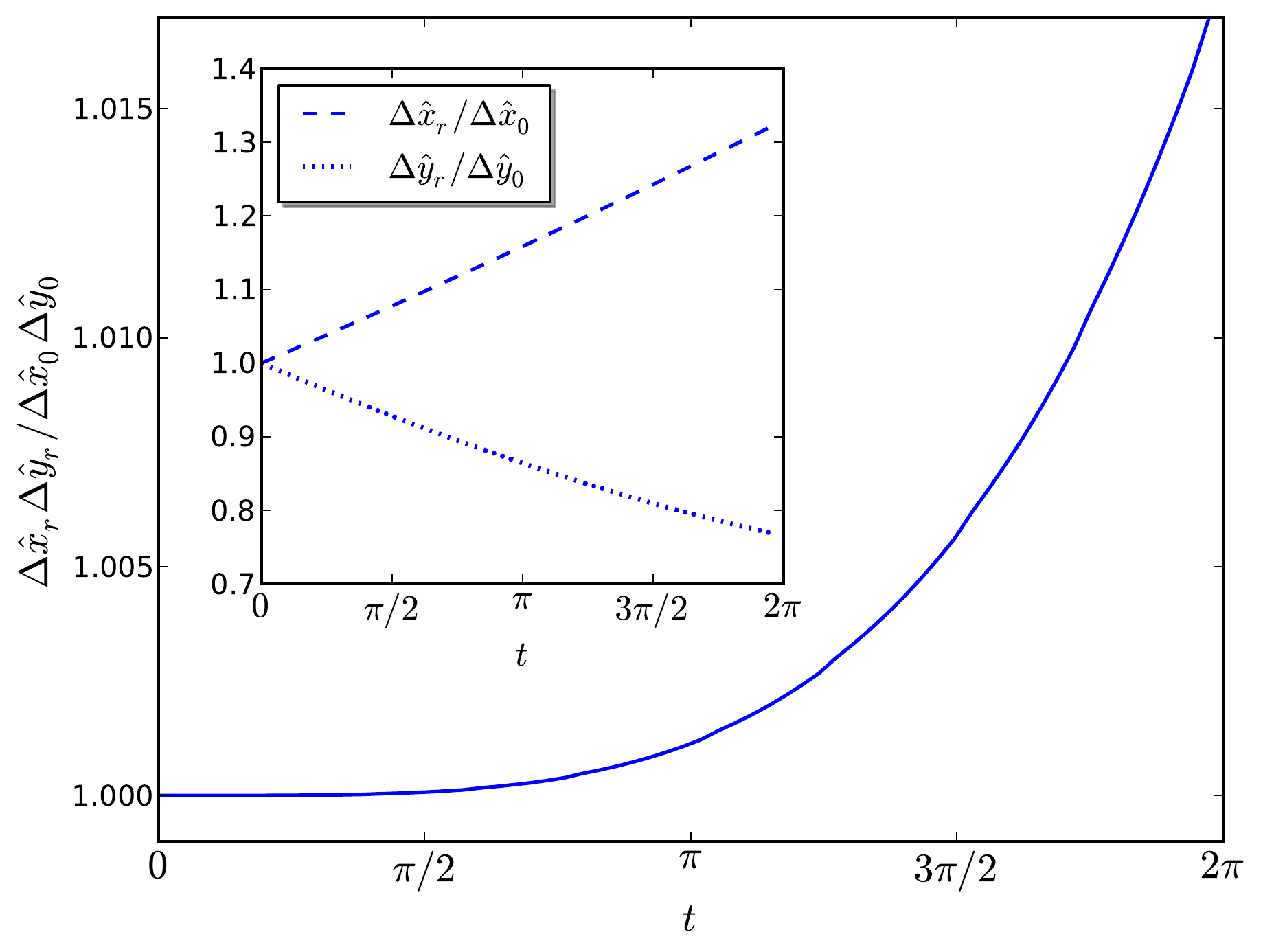}
  \caption{(color online) The main plot shows the normalized uncertainty relation $\Delta \hat{x}_r\Delta \hat{y}_r/\Delta \hat{x}_0\Delta \hat{y}_0$ for the first cycle in the weak coupling regime. The sudden increasing in time shows the short period in which the squeezing remains valid. The subplot shows the individual normalized variance, the quadrature $x_r$ ($y_r$) becomes linearly increasing (decreasing).}\label{squeezing}
\end{figure}
\end{center}

We plot in Fig.~\ref{wigner2D_small_k_delta}-a the Wigner function for the NLO. For short times, we see that due to the weak coupling the two components of the Wigner function associated with the qubit eigenstates are superposed (\textit{i.e.} $\ket{\tilde{\alpha}_\uparrow} \approx \ket{\tilde{\alpha}_\downarrow}$). As anticipated, in contrast with the linear case we can see that the presence of the additional Kerr-like term gives rise to non-classical features. In particular Fig.~\ref{wigner2D_small_k_delta}-a  shows the emergence of quadrature squeezing, with squeezing axes that rotate clockwise in the $xy$-plane. Defining two arbitrary canonical quadratures ($\phi$ is the angle of rotation measured from the $x$-axis to $x_r$-axis)
\begin{eqnarray}
 \hat{x}_r &=& \frac{1}{2}\left(\hat{a}e^{-i\phi} + \hat{a}^\dag e^{i\phi} \right),\\
 \hat{y}_r &=& \frac{1}{2i}\left(\hat{a}e^{-i\phi} - \hat{a}^\dag e^{i\phi} \right).\label{phi}
\end{eqnarray}
we numerically find for each time $t$ the angle $\phi$ that minimize the uncertainty of $ \Delta \hat{y}_r$ (where $\Delta \hat{O} = \langle \hat{O}^2 \rangle - \langle \hat{O} \rangle^2$). The results are given in Fig.~\ref{squeezing} and quantitatively demonstrate the presence of squeezing for short times (the results are normalized with respect to the coherent state uncertainty $\Delta \hat{x}_0 = \Delta \hat{y}_0 = 1/2$).

Another interesting feature is that whereas for short times the Wigner function remains positive, for longer times it assumes negative values. Interestingly, the appearance of relevant negative regions corresponds to the stabilization zone of the negativity --- for example at $t = 50\pi$ (see Fig.~\ref{wigner2D_small_k_delta}-b). 

\subsection{Strong qubit-NLO coupling regime : Numerical solution for k $\approx$ 1, $\delta \ll 1$.}\label{StrongCoupling}
In this section, we solve numerically the full dynamics involving the Hamiltonian in Eq.~(\ref{han}) without restricting to the weak coupling regime. In order to do that, we have expanded the oscillator state in the Fock basis, properly truncated to obtain a sufficient numerical accuracy. 

\begin{figure}
\includegraphics[scale = 0.45]{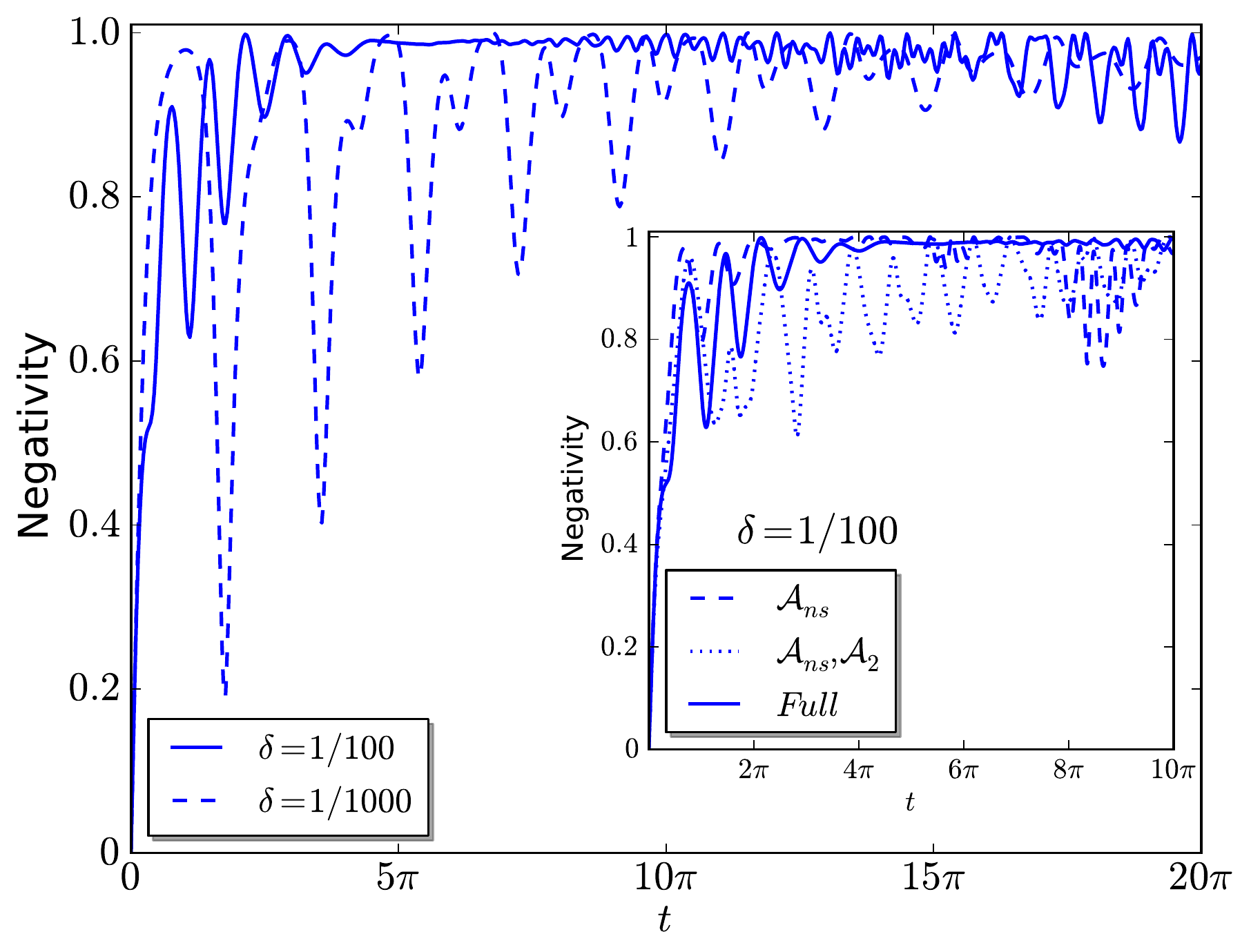}
\caption{(color online) In the main plot we show the numerical results for the negativity for $k = 0.5, \alpha = 2$ and varying $\delta$. In contrast with $\{k, \delta\} \ll 1$, here we have achieved a higher entanglement as well as a faster stabilization zone. In the subplot we compare the entanglement generated for $k = 0.5, \alpha = 2, \delta = 1/100$ using approximated Hamiltonian, the solid line is for a full Hamiltonian without approximation. The dashed line consider only number states in the quartic potential, and finally the dotted line consider up to four-phonon transitions in the quartic potential.}\label{N_nonzero}
\end{figure}

Regarding the generation of entanglement between the qubit and the NLO, the effects of a strong coupling are that the two main features that we have individuated in the previous section are further enhanced. First, the entanglement negativity reaches higher values with respect to the absence of nonlinearities. Second, the entanglement reaches the stabilization region faster then for the weak coupling regime.  As an example, we have plotted in Fig.~\ref{N_nonzero} the negativity for $k = 0.5, \alpha = 2$, and two values for $\delta = \{1/100, 1/1000\}$. We can see that the negativity stabilizes already for $t\approx 5\pi$ ($\delta = 1/100$) close to the maximal reachable value of $1$. This stability is sustained quite well in a window of time from $t = 5\pi$ to $t = 10\pi$, after which it starts to oscillate. Remarkably, in this regime the collapse and revival dynamics is entirely absent. The combination of a high amount of entanglement and the suppression of negativity oscillations provides a long time 
window in which the entanglement is maximal or near-maximal, in strong  contrast to the linear case where maximal negativity is achieved only at defined times (odd multiples of $t=\pi$). Here timing selection is no longer a concern in order to achieve high negativity, representing in turn a relevant practical advantage.

Furthermore, it is important to note in the subplot in Fig.~\ref{N_nonzero}, that we have also considered the contribution of approximated Hamiltonian regarding to only number-state contribution and up 
to two-phonon transitions in the dynamics (see Eq.~\ref{a4}). As we can see in solid line, the full dynamics ---i.e including up to four-phonon transitions--- provides an entanglement plateau in time domain better than the other approximated cases.

\begin{center}
\begin{figure}
\includegraphics[width = 0.95\textwidth, height = 0.75\textheight]{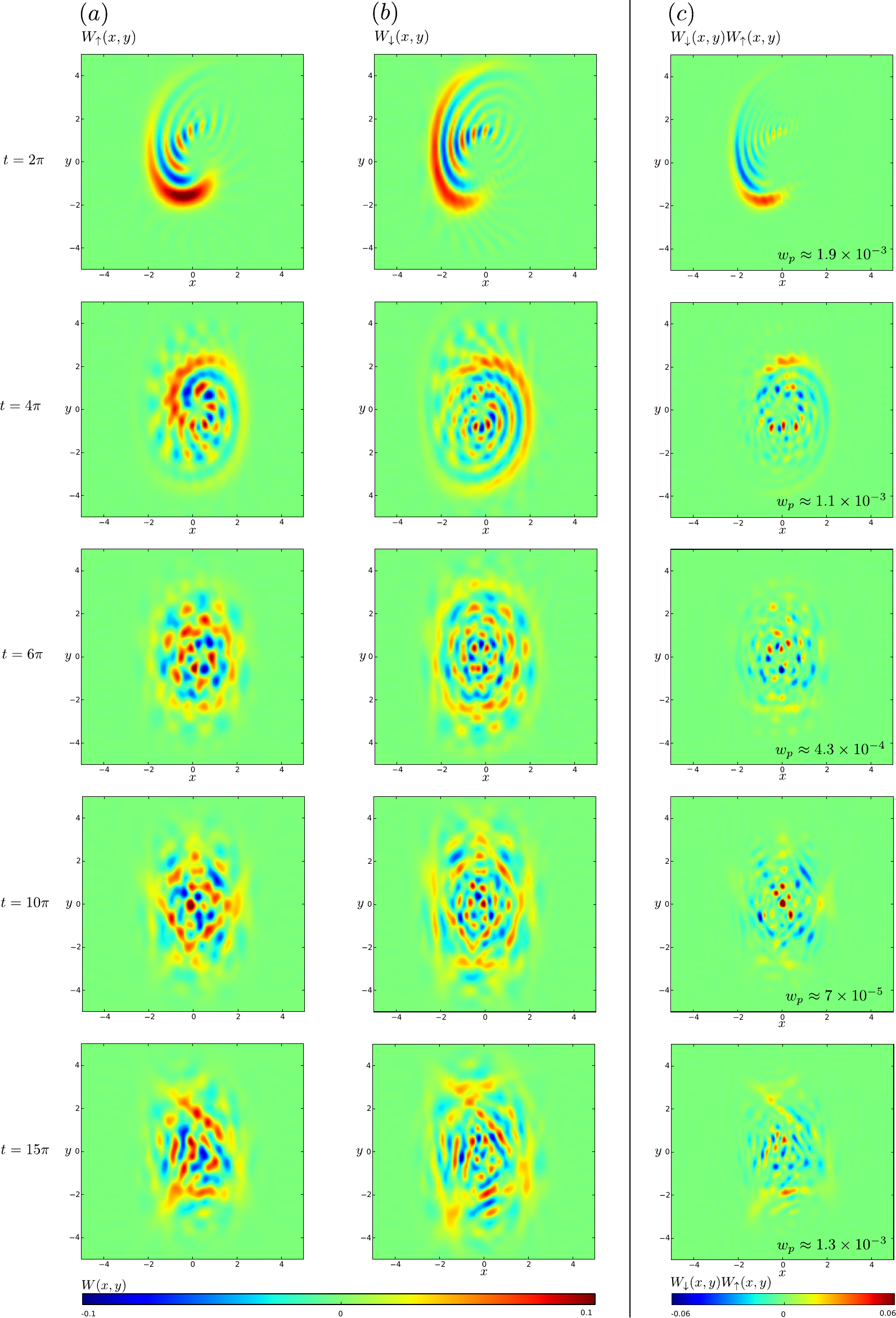}
\caption{(color online) Here we provide a pictorial explanation for the entanglement enhancement for $k = 0.5, \alpha = 2, \delta = 1/100$ at different times $t = 2\pi, 4\pi, 6\pi, 10\pi, 15\pi$. In Figs.~(a) and (b), we plot the Wigner function for the oscillator state for each spin component $W_{\uparrow, \downarrow}(x,y)$. In column (c), we show the product between $W_{\uparrow}(x,y)$ and $W_{\downarrow}(x,y)$. The number $w_p$ in the corner corresponds to the integration of the product over all the $xy-$phase space (Eq.~\ref{wp}). The small overlap between $W_{\uparrow}(x,y)$ and $W_{\downarrow}(x,y)$ then shows that the states corresponding to the latter are quasi-orthogonal, thus allowing for the establishment of maximal entanglement.}\label{Wig_005}
\end{figure}
\end{center}

In order to better understand the enhancement of the qubit-NLO entanglement we have calculated the Wigner function of the oscillator state conditioned to the two qubit eigenstates $i.e., \hat{\rho}(t)_{\uparrow}^{osc} = \bra{\uparrow}\hat{\rho}(t)\ket{\uparrow}$ (or $ \hat{\rho}(t)_{\downarrow}^{osc} = \bra{\downarrow}\hat{\rho}(t)\ket{\downarrow}$). In Fig.~\ref{Wig_005}, we have plot $W_{\uparrow, \downarrow}(x,y) = \frac{1}{\pi\hbar}\int_{-\infty}^\infty\left<x + x'\left| \hat{\rho}_{\uparrow, \downarrow}^{osc}\right| x-x' \right>e^{-2iyx'/\hbar}dx'$ at $t = 2\pi, 4\pi, 6\pi, 10\pi, 15\pi$ for each qubit component, together with their product. We can see that the overlap between the two functions sensibly decades already after the first cycle ($t = 2\pi$). In order to show this quantitatively we illustrates in Fig.~\ref{Wig_005}-c the overlap of the product $W_{\uparrow}(x,y)W_{\downarrow}(x,y)$ together with its integration over all $xy-$phase space
\begin{equation}
 w_p = \int_{-\infty}^{+\infty}\int_{-\infty}^{+\infty}W_{\uparrow}(x',y')W_{\downarrow}(x',y')dx'dy'.\label{wp}
\end{equation}

\onecolumngrid
\twocolumngrid

In other words, this shows that the conditioned Wigner functions $W_{\uparrow, \downarrow}(x,y)$ correspond to two almost orthogonal states which implies that maximally entanglement can be established between the qubit and the oscillator. The quasi-orthogonality is quantified using Eq.~(\ref{wp}) and shown in Fig.~\ref{Wig_005}-c.

\begin{figure}
\centering
\includegraphics[scale = 0.45]{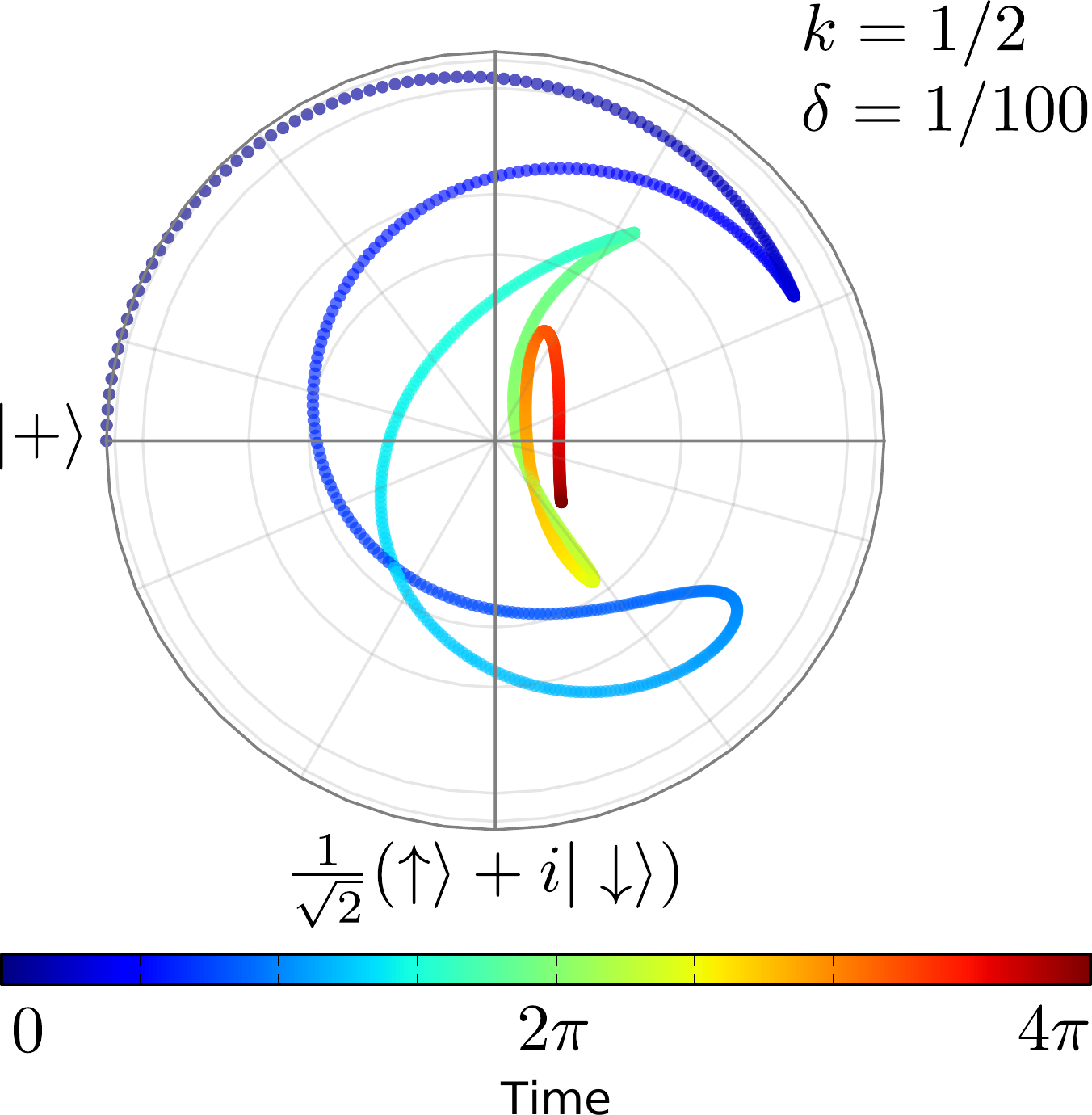}
\caption{(color online) We illustrate the reduced density qubit operator in Bloch Sphere (top-view) for two cycles $0 \leq t \leq 4\pi$. The qubit shows a strong precession in the dynamics.}\label{Bloch_Sphere_2}
\end{figure}

As before, we also calculated numerically the reduced density matrix for the qubit. In the presence of nonlinear coupling, the qubit exhibits an open cycle whose precession depends on the strength of the nonlinearity. For $\delta \ll 1/1000$ the reduced qubit evolution tends to the quantum harmonic potential case, and therefore each cycle is closed. On the other hand, as $\delta$ increases, the qubit reaches a stationary point at times comparable to the entanglement stabilization region (see Fig. \ref{Bloch_Sphere_2}).

\begin{center}
\begin{figure}
\includegraphics[scale = 0.45]{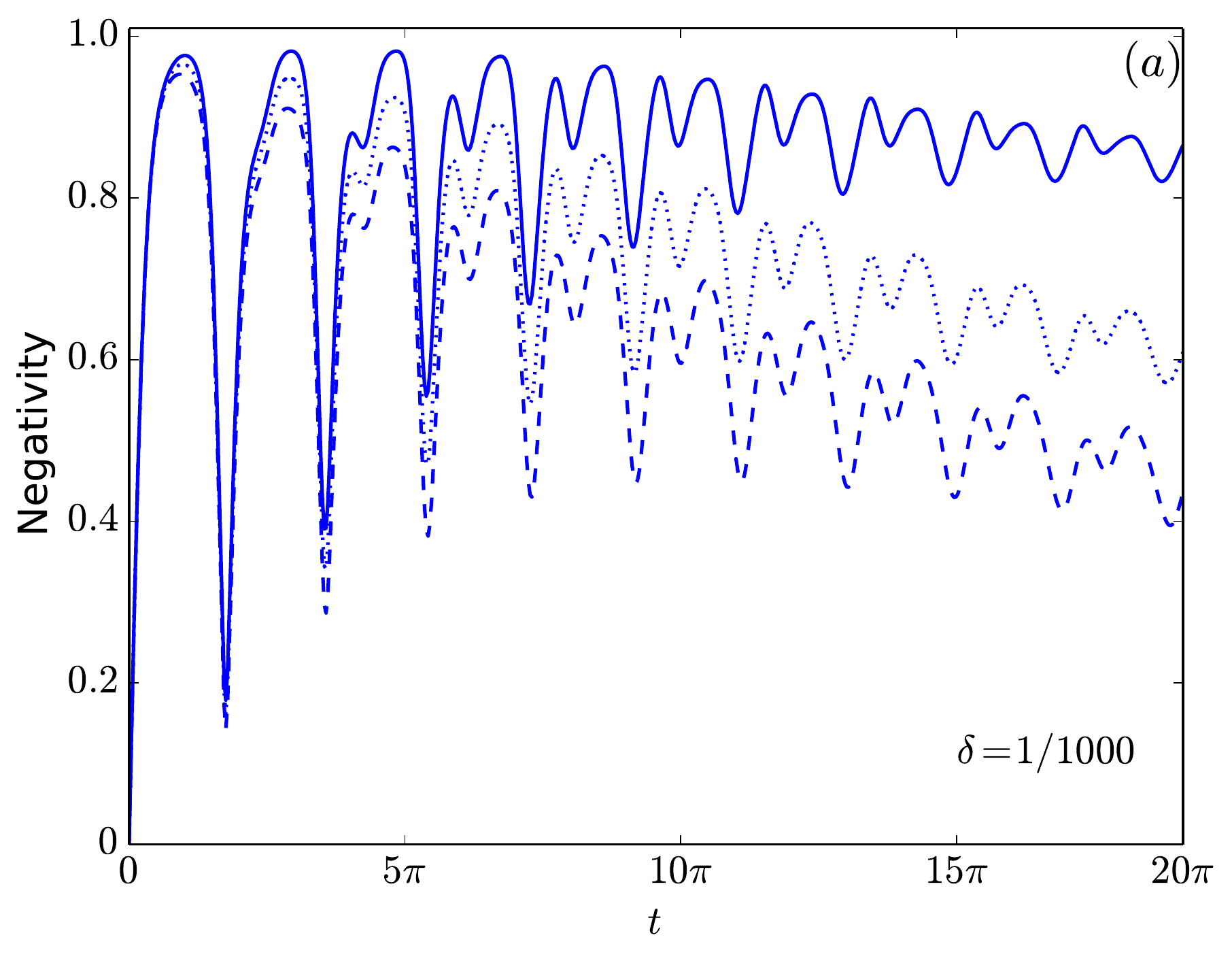}
\includegraphics[scale = 0.45]{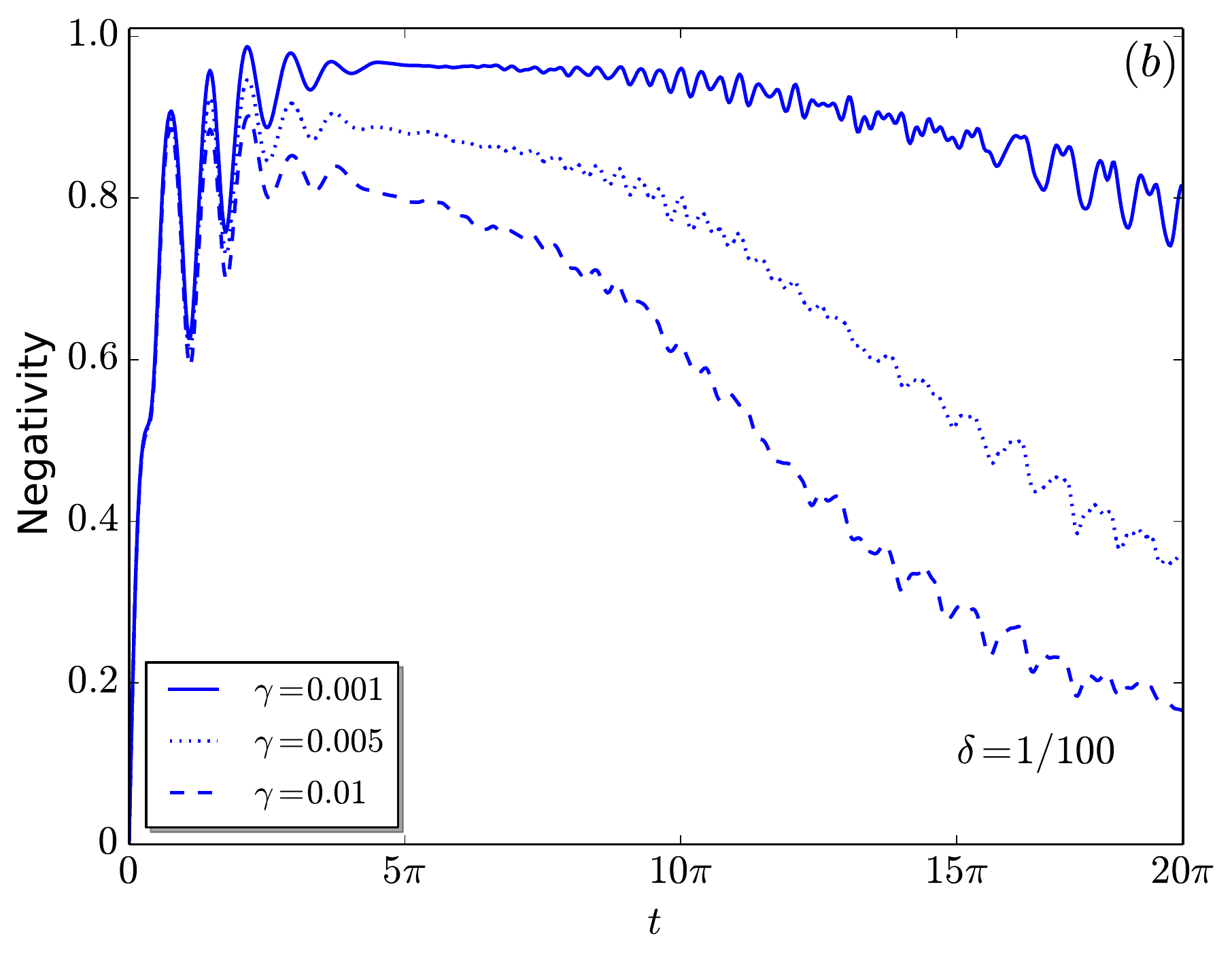}
\caption{(color online) negativity for the open quantum system for different values of the dissipation ratio $\gamma$, and different values of $\delta$. Here, $k = 0.5, \alpha = 2$.}\label{Master}
\end{figure}
\end{center}

Finally, we considered the detrimental effects of noise in the dynamics of the NLO. We modeled the system with the following master equation in Lindblad form at zero temperature 
\begin{eqnarray}
 \nonumber \dot{\hat{\rho}}(t) &=& -i[\hat{H}_{s},\hat{\rho}(t)] + \frac{\gamma}{2}(2\hat{a}\hat{\rho}(t)\hat{a}^\dag - \hat{a}^\dag\hat{a}\hat{\rho}(t) - \hat{\rho}(t)\hat{a}^\dag\hat{a})\\
\end{eqnarray}

where $\gamma$ is the oscillator damping rate. In Fig.~\ref{Master} we show the main effects of the losses. We can see that in the strong coupling regime the presence of the environment degrades the  qubit-NLO entanglement but the main features observed in the previous sections are still present. In particular, both the enhancement of entanglement with respect to the linear case and the entanglement stabilization are robust for small losses.

We have already mentioned some of the primary experimental setups in the introduction, let us now briefly examine some of those. The strong-coupling regime can be achieved using a qubit encoded in an electron on a quantum dot or a Cooper pair on a small superconducting island, coupled to an oscillator consisting of a vibrating gate electrode. In Ref. \cite{ARMOUR} the authors consider a micromechanical resonator capacitively coupled to a Cooper pair box (CPB). Here, they can reach substantial coupling in the range of $g = 5 - 50$ MHz, and with the current technology $k \approx 1$ can be performed. Following with this regime, a qubit can also be modeled in clockwise and anti clockwise circulating currents in a superconducting loop. For instance, in Ref. \cite{RABL} the authors accomplish a strong coupling between a single electronic spin qubit associated with a nitrogen-vacancy impurity in diamond and the quantized motion of a magnetized nanomechanical resonator tip. Here, the 
dimensionless coupling is approximately $k \approx 0.1$ (For further details related with the full set of parameters see Ref. \cite{RABL}). On the other hand, the weak coupling regime can be realized in systems where a quantum dot is coupled to a mechanical oscillator, where this resonator is modulated by changing the local lattice of the host material \cite{KOLKIRAN}.

Another candidate setting for the implementation is given by trapped ions, where the strong coupling between hyperfine internal states of an ion and its motional degree of freedom has been shown in a variety of configurations \cite{IONS}. Moreover, the ion internal state can also be coupled to a cantilever under realistic conditions, for example, for a doubly clamped cantilever frequency of 19.7 MHz \cite{HAYE}. The coupling strength ---which can be switched on and off--- for a cadmium ion is given by $g \approx 52.5$ kHz \cite{HENSINGER1}, and therefore $k \approx 10^{-3}$.

Concerning possible implementations of non-linear quantum oscillators, various experimental platforms can be envisaged. As said, trapped ions can host qubit-oscillator systems. These platforms can also implement non-linear oscillators and in fact, by using a tunable set of parameters, the authors of Ref.~\cite{HENSINGER} showed how to encompass both linear and nonlinear potentials (anharmonic and double-well)---in order to achieve the efficient separation and re-combination of ions in surface ion-trap geometries using effective potentials. Furthermore, nonlinearities can be generated as a result of static and longitudinal compressive force in suspended nanomechanical beams \cite{KOLKIRAN}. For instance, for values of the length ($L$), thickness ($d$), and width ($w$) of the nanomechanical beam in the range of $L \approx 200 - 400$nm, $d \approx 5-10$nm, and $w \approx 10-20$nm  a nonlinear strength of the order of $\delta \approx 10^{-2}$ can be obtained. Finally, nonlinearities can be achieved in a 
mechanical oscillator in the form of a nano-cantilever cooled to its ground state. There, a ferromagnetic impurity in the cantilever tip (nano-magnets) can induce non-linear potentials via high homogeneous external magnetic fields in Helmholtz coil configuration \cite{ANDERSSON} (for an overview of quantum mechanical systems see Ref. \cite{POOT}). All the mentioned systems are promising candidates in order to achieve the nonlinearity we have considered in this work. In combination with the qubit-oscillator coupling, these schemes points at the actual possibility of implementing the qubit-NLO coupling, being the non-linearity the most challenging task to achieve in an experiment.

\section{Concluding Remarks}\label{Concluding Remarks}
We have investigated a qubit (spin) coupled to a quartic nonlinear oscillator through a conditional displacement Hamiltonian. The dynamics begins from a separable initial state composed of a qubit superposition state $(\ket{\uparrow} + \ket{\downarrow})/\sqrt2$ and a coherent state $\ket{\alpha}$ for the oscillator. Throughout the paper we have used two relevant parameters, namely, the qubit-NLO coupling $k$ and the nonlinearity $\delta$. We first recalled the results for the case $\delta = 0$. Here, the entanglement generation is due to the superposition principle of the hybrid system and it shows a periodic dynamics. On the other hand, when $\delta \neq 0$ and in the \textit{weak coupling} regime we analytically show that a new Kerr-like term appears in the dynamics leading to \textit{i)} quadrature squeezing of the oscillator state, \textit{ii)} the suppression of the entanglement decay by the appearance of a stabilization region, and \textit{iii)} an enhancement of the entanglement negativity compared to 
the linear case of $\delta = 0$. 

The most interesting case corresponds to the \textit{strong coupling regime}, when we see that two- and four-phonon transitions play a relevant role both in the entanglement stabilization and in its enhancement. In particular, the entanglement negativity can reach its maximal value by virtue of the orthogonalization of the oscillator states relevant to the present dynamics. Furthermore, solving numerically the corresponding master equation, we have shown that these effects remain robust to the presence of decoherence in the oscillator system.

Finally, we have considered in some details different possible experimental implementations for each regime considered here. Witnessing this type of hybrid entanglement is a hard task, however following the protocol in Ref. \cite{PARK} we can give a full proof of the violation of a Bell inequality for $\delta = 0$ (and for the weak coupling regime when $\delta \approx 10^{-3}$). Nevertheless, a full benchmark in the strong coupling regime remains unsolved and will be subject of future work.

\bigskip
%\acknowledgments
VM is supported by the Comisi\'on Nacional de Investigaci\'on Cient\'ifica y Tecnol\'ogica (CONICYT - Becas Chile ID 72110207). AF acknowledges funding from John Templeton Foundation (grant ID 43467). SB acknowledges the EPSRC grant EP/J014664/1.

\appendix
\section{Wave function in absence of nonlinearities}\label{appA}
 In order to obtain the time evolution operator for this case, we use a direct consequence of the similarity transformation which holds the following
\begin{equation}
 \hat{T}f\left(\{\hat{X}_i\} \right)\hat{T}^\dag = f\left(\{\hat{T}\hat{X}_i \hat{T}^\dag\}\right),
\end{equation}

the above equation is satisfied for any function $f$, unitary operator $\hat{T}$, and arbitrary set of operators $\{\hat{X}_i\}$. Hence we take in particular
\begin{eqnarray}
 \hat{T} &=& e^{-k\hat{\sigma}_z(\hat{a}^\dag - \hat{a})},\\
 f\left(\{\hat{X}_i\}\right) &=& \hat{U}(t) = e^{-it\hat{H}}
\end{eqnarray}

here $\{\hat{X}_i\} = \{\hat{a},\hat{\sigma}_z\}$. Using the Baker-Campbell-Hausdorff (BCH) relation, it is straightforward to show the following transformations
\begin{eqnarray}
 \hat{T}\hat{a}\hat{T}^\dag &=& \hat{a} + k\hat{\sigma}_z,\\
 \hat{T}\hat{\sigma}_z\hat{T}^\dag &=& \hat{\sigma}_z.
\end{eqnarray}

Using both the Similarity Transformation as well as the BCH relation, it is easy to obtain the analytical expression for the time evolution operator
\begin{eqnarray}
\nonumber \hat{U}(t) &=& \mathrm{exp}\left[ik^2(t-\mathrm{sin}(t))\right]\\
&\times&\mathrm{exp}\left[k\hat{\sigma}_z(\eta\hat{a}^{\dag} - \eta^*\hat{a})\right]\mathrm{exp}\left[-i\hat{a}^{\dag}\hat{a}t\right]
\end{eqnarray} 

where,
\begin{equation}
\eta = 1-\mathrm{exp}\left[-it\right].
\end{equation}

Therefore, the time evolution for the initial state (Eq. (\ref{psi0})) corresponds to
\begin{eqnarray}
\nonumber \ket{\psi(t)} &=& \frac{1}{\sqrt{2}}\ket{\uparrow}\otimes\hat{D}(k\eta)\hat{D}(\alpha e^{-it})\ket{0}\\
&+&\frac{1}{\sqrt{2}}\ket{\downarrow}\otimes\hat{D}(-k\eta)\hat{D}(\alpha e^{-it})\ket{0},
\end{eqnarray}

taking into account that $\hat{D}(\alpha_1)\hat{D}(\alpha_2) = \mathrm{exp}\left[(\alpha_1\alpha_2^* - \alpha_1^*\alpha_2)/2\right]\hat{D}(\alpha_1 + \alpha_2)$, we can finally obtain the final form shown in Eq. (\ref{negdelta0}).

\section{Wave function in the weak qubit-NLO coupling regime}\label{appB}
In order to obtain the unitary operator for the RWA Hamiltonian in Eq.~(\ref{ah_app}) we will use the same techniques as before, let's consider the following approximation 
\begin{equation}
\hat{T}\hat{U}(t)_{app}\hat{T}^\dag \approx \mathrm{exp}[-it(1+6\delta)\hat{a}^\dag\hat{a} - 6i\delta t(\hat{a}^\dag\hat{a})^2],\label{tuappt}
\end{equation} 

where we have neglected products proportional to $\{k\delta,k^2\delta,k^3\delta\} \ll 1$. Taking into account that
\begin{eqnarray}
\nonumber &&\mathrm{exp}[-it(1+6\delta)\hat{a}^\dag\hat{a}]\hat{T} = \\
\nonumber &&\mathrm{exp}[-k\hat{\sigma}_z(\hat{a}^\dag e^{-i(1+6\delta)t}-\hat{a}e^{i(1+6\delta)t})]\mathrm{exp}[-i(1+6\delta)t\hat{a}^\dag\hat{a}].\\
\end{eqnarray} 

Multiplying on the left by $\hat{T}^\dag$ and on the right by $\hat{T}$ the Eq. (\ref{tuappt}), we can finally obtain the time evolution operator
\begin{eqnarray}
\nonumber \hat{U}_{app} &=& \mathrm{exp}[-k\hat{\sigma}_z(\hat{a} - \hat{a}^\dag)\mathrm{exp}[-6\delta it(\hat{a}^\dag\hat{a})^2]\\
\nonumber &\times& \mathrm{exp}[-k\hat{\sigma}_z(\hat{a}^\dag e^{-i(1+6\delta)t} - \hat{a}e^{i(1+6\delta)t})]\\
&\times& \mathrm{exp}[-i(1+6\delta)t\hat{a}^\dag\hat{a}].
\end{eqnarray}

Using the above, it is straightforward obtain the wave function for this case
\begin{eqnarray}
 \nonumber \ket{\psi(t)} &=& \frac{1}{\sqrt{2}}\mathrm{exp}\left[k(\hat{a}^\dag - \hat{a})\right]\mathrm{exp}\left[-6it\delta(\hat{a}^\dag\hat{a})^2\right]\\
 \nonumber &\times&\mathrm{exp}\left[\frac{k}{2}(\alpha - \alpha^*)\right]\ket{\uparrow}\otimes\ket{e^{-i(1+6\delta)t}(\alpha - k)}\\
 \nonumber &+& \frac{1}{\sqrt{2}}\mathrm{exp}\left[k(\hat{a} - \hat{a}^\dag)\right]\mathrm{exp}\left[-6it\delta(\hat{a}^\dag\hat{a})^2\right]\\
 \nonumber &\times&\mathrm{exp}\left[\frac{k}{2}(\alpha^* - \alpha)\right]\ket{\downarrow}\otimes\ket{e^{-i(1+6\delta)t}(\alpha + k)}\\ \label{psi_app}
\end{eqnarray}

In general we have taken a real amplitude for the coherent state ($\alpha = 2$), hence the phase appearing in Eq. (\ref{psi_app}) vanishes giving us the Eq.~(\ref{asmall}).

\end{document}